\documentstyle[aaspp4,emulateapj5,apjfonts]{article}

\gdef\hdfn{HDF-N}

\gdef\h50min{$h_{50}^{-1}$}
\gdef\kms{km\,s$^{-1}$}
\gdef\hk{Ca\,{\sc ii} H and K}

\gdef\3727{[O\,{\sc ii}]\,3727\,\AA}
\gdef\5007{[O\,{\sc iii}]\,5007\,\AA}
\gdef\oii{[O\,{\sc ii}]}
\gdef\4ang{4000\,\AA}
\lefthead{van Dokkum \& Ellis}
\righthead{Early-Type Galaxies in HDF-North}
\slugcomment{Accepted for publication in Astrophysical Journal Letters}
\begin{document}

\title{On the Assembly History of Early-Type Galaxies in the Hubble
Deep Field North\altaffilmark{1,2}}
\author{Pieter~G.~van Dokkum\altaffilmark{3,4} and
Richard~S.~Ellis\altaffilmark{4}
}

\altaffiltext{1}{Based on observations with the NASA/ESA {\em
Hubble Space Telescope}, obtained at the Space Telescope Science
Institute, which is operated by AURA, Inc., under NASA contract
NAS 5--26555.} \altaffiltext{2} {Based on observations obtained at
the W.\ M.\ Keck Observatory, which is operated jointly by the
California Institute of Technology and the University of
California.} \altaffiltext{3}{Department of Astronomy, Yale
University, New Haven, CT 06520-8101} \altaffiltext{4}{California
Institute of Technology, MS105-24, Pasadena, CA 91125}

\begin{abstract}

We present deep Keck spectroscopy for a sample of $I_{814}< 22.5$
field early-type galaxies selected morphologically in the redshift
range $0.56\leq z \leq 1.02$ in the Hubble Deep Field North
(HDF-N). Using velocity dispersions determined from the Keck
spectra in conjunction with structural parameters measured from
the deep WFPC2 images we study the evolution of the $M/L_B$ ratio
and the fundamental plane (FP) with redshift. For the majority of
galaxies the trends observed are very similar to those determined
earlier for rich clusters. The systematic offset between HDF-N galaxies
and cluster galaxies is
$\Delta \ln M/L_B = -0.14 \pm 0.13$, corresponding to an 
age difference of only $16\pm 15$\,\% at
$\overline{z} =0.88$. However,
we find enhanced H$\delta$ absorption
of equivalent width $4.0^{+0.9}_{-0.5}$\,\AA\
in the mean spectrum of the ten galaxies, indicating the presence
of young stars. We infer that the galaxies
have composite stellar populations, consisting of a low mass
young component in addition to a 
dominating old component.
As the bulk of the stellar mass must have formed at
$z\gtrsim 2$ our results argue against formation scenarios
involving major
mergers of gas-rich disk systems 
at $1\lesssim z\lesssim 1.5$, and we conclude that
$z\simeq 1$ early-type galaxies
were either assembled at higher redshift or in mergers involving little gas.
The ubiquitous enhanced Balmer lines and
the presence of tidal features in two of the galaxies
lend some support to the latter hypothesis.
The main uncertainty in the analysis is the small
sample; larger samples are needed to study the interplay between
the evolution of stellar populations and morphology in detail.

\end{abstract}


\section{Introduction}

As continued merging is a central prediction of hierarchical
assembly models, much effort has been expended in attempting to
detect a decline with redshift in the abundance of well-formed
early-type galaxies (e.g., {Im} {et~al.} 2002, Bell et al.\ 2003,
Treu 2003). Present results are somewhat inconclusive, largely
because of the lack of high resolution imaging data for
sufficiently large samples, the difficulty disentangling
luminosity evolution and changes in the number density, and the
effects of large scale structure. The most comprehensive study to
date is the analysis of a very large color-selected sample by Bell
et al.\ (2003), who suggest that the stellar mass in red galaxies
may have increased by a factor 2--3 between $z=1$ and $z=0$.

Two alternative routes to the formation history of field
spheroidals have emerged in the past few years. {Treu} {et~al.} (1999),
{van Dokkum} {et~al.} (2001) and others have attempted to constrain
evolution in the fundamental plane (FP) differentially with
respect to that observed in clusters. Differential tests are less
sensitive to selection effects and provide valuable constraints on
environmental trends expected in hierarchical models
(e.g., {Kauffmann} 1996). Furthermore, the FP gives
information on the mass-to-light ($M/L$) ratios of galaxies, which
can be used to interpret the observed evolution of the luminosity
function in terms of the underlying mass function. Again, current
results are not yet conclusive. Field and cluster samples with
$z<0.5$ have similar $M/L$ ratios
({Kochanek} {et~al.} 2000; {van Dokkum} {et~al.} 2001; {Treu} {et~al.} 2001), but there are
indications that differences set in at higher redshift
({Treu} {et~al.} 2002).

A second development follows the discovery by {Menanteau}, {Abraham}, \&  {Ellis} (2001)
that $\simeq$25\,\% of field early-type galaxies in the Hubble Deep
Fields (HDFs) show color inhomogeneities such as blue cores,
quantified by non-uniformity measures across the resolved HST image.
Such abnormalities may indicate recent (merger-induced) star formation
and, if so, their study offers a possible route to an {\em in situ}
mass assembly rate ({Benson}, {Ellis}, \& {Menanteau} 2002). A key question therefore is
the extent to which spectroscopic diagnostics of recent activity
are seen in distant early-type galaxies (e.g., {Barger} {et~al.} 1996).

In this {\em Letter}, we address the spectroscopic properties of a
representative sample of HST-selected early-type galaxies in the
\hdfn, with a median redshift $z\simeq$0.9. First we investigate
whether the $M/L$ ratios of field and cluster early-type galaxies
diverge beyond $z\sim 0.5$, as suggested by the work of
{Treu} {et~al.} (2002). Secondly, we consider the evidence for recent
star formation and mergers using spectroscopic diagnostics. As the
\hdfn\ represents a small sample, we consider this an initial
exploration of these issues, demonstrating the value of more
ambitious surveys now possible with wide field spectrographs such
as DEIMOS (Faber et al 2002). We assume $\Omega_m=0.3$ and
$\Omega_{\Lambda}=0.7$ throughout.

\begin{table*}[t]
\centering
\caption{Early Type Galaxies in HDF-N Proper}
\begin{tabular}{ccccccccccc}
\hline
\hline
ID & HDFN\_J & $z$ & $I^{\rm tot}_{814}$ & S/N
& $\sigma$ & error & $\log r_e$ & $\mu_e$ & H$\delta$ & [O II] \\
   & &    & (Vega)   &    &  (km/s)   & & (arcsec)&
($I_{814}$) & (\AA) & (\AA)  \\
\hline
1 & 123640.02+621207.4 & 1.017 & 21.84 & 14  & 235 & 46 & $-0.680$ & 21.79 & 5.4 & 1.7 \\
2 & 123643.16+621242.2 & 0.847 & 21.28 & 23   & 225 & 32 & $-0.603$ & 21.44 & 1.7 & 1.1 \\
4 & 123656.65+621220.2 & 0.954 & 22.09 & 13  & 174 & 32 & $-0.490$ & 22.91 & 2.6 & 2.4 \\
5 & 123643.63+621218.3 & 0.751 & 22.26 & 11   &  --   &  --  &  --  & --   & 7.0 & -- \\
7 & 123646.14+621246.5 & 0.904 & 21.89 & 12   & 244 & 48 & $-0.635$ & 22.14 & 4.6 & $<1.0$\\
8 & 123650.27+621245.8 & 0.680 & 20.96 & 21   & 175 & 19 & $-0.483$ & 21.95 & 3.3 & -- \\
13 & 123644.38+621133.2 & 1.013 & 21.06 & 20  & 286 & 29 & $-0.220$ & 23.06 & 1.3 & $<1.0$ \\
14 & 123700.56+621234.7 & 0.562 & 20.94 & 23  & 114 & 19 & $-0.359$ & 22.30 & 4.7 & -- \\
16 & 123643.81+621142.9 & 0.765 & 20.43 & 18  & 161 & 27 & $-0.423$ & 21.68 & 1.3 & 3.0 \\
17 & 123646.35+621404.7 & 0.960 & 20.84 & 26  & 182 & 29 & $-0.897$ & 20.15 & 2.9 & 3.9 \\
\hline

\end{tabular}
\end{table*}

\section{Data}

\subsection{Sample Selection and Spectroscopy}

Early-type galaxies in \hdfn\ were selected morphologically from
the catalog of {Ellis}, {Abraham}, \&  {Dickinson} (2001).
The reliability of the classifications is discussed
in some detail by {Menanteau} {et~al.} (1999). Our present sample
comprises a subset with $I_{814} < 22.5$ and $z>0.5$.
Spectroscopic redshifts are available for almost all of
the \hdfn\ sources to our limit from {Cohen} {et~al.} (2000).


Spectroscopic observations were conducted at the Keck I telescope
using the Low Resolution Imaging Spectrograph ({Oke} {et~al.} 1995)
on 2000 March 29--30 and 2001 March 27--28. A 7 hour exposure in
mediocre seeing was obtained in 2000 using the 600 lines mm$^{-1}$
grating blazed at 7500\,\AA.  Superior data were obtained in 2001
in good conditions, using the 600 lines mm$^{-1}$ grating blazed
at $1\,\mu$m. Between exposures the galaxies were moved along the
slit to facilitate sky subtraction. 
The latter 12 hour exposures provided the bulk of the data
comprising 10 targets. The instrumental resolution $\sigma_{\rm
instr}\approx 80$\,\kms\ at 8000\,\AA. The spectral coverage is
approximately 7300--9900\,\AA. The data were reduced
following standard procedures for dithered multi-slit data; the
reduction of a very similar dataset is explained in detail in
{van Dokkum} \& {Stanford} (2003).

Internal kinematics were determined by fitting broadened template
star spectra to the galaxy spectra. Fits were performed in real
space, since S/N weighting and masking of spectral features is
more troublesome in Fourier space (see, e.g., {Kelson} {et~al.} 2000a for
details). The template stars cover the wavelength
range $3200 - 5200$\,\AA\ at high resolution
(see {van Dokkum} \& {Stanford} 2003). Velocity dispersions were derived for
9/10 targets (Table~1). The uncertainties do not include a
systematic error of $\approx 8$\,\% as determined from varying the
fitting region, template stars, and the continuum filtering. The
S/N in the spectra ranges from 12 to 26 per \AA\ . Studies at low
redshift have shown that for S/N$\lesssim 15$ systematic effects
begin to dominate the errors (e.g., {J\o{}rgensen}, {Franx}, \&  {Kj\ae{}rgaard} 1995), and we
find that both the S/N and the formal errors indicate that the
exposure time was just sufficient for measuring velocity
dispersions of these faint galaxies.

\begin{figure*}[t]
\begin{center}
\leavevmode
\hbox{%
\epsfxsize=14.2cm
\epsffile{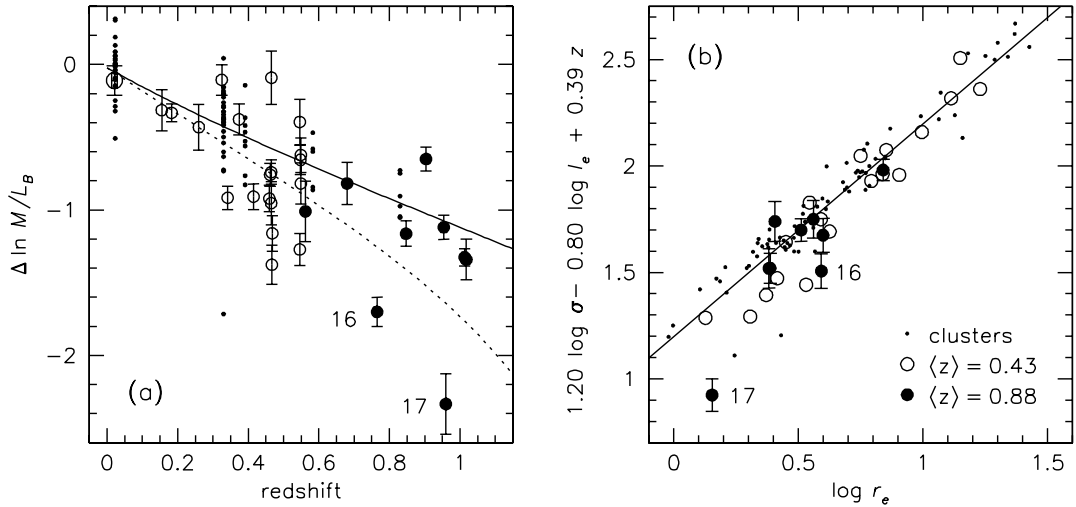}}
\figcaption{
\small
Panel {\em (a)} shows the $\Delta \ln M/L_B$ ratios of
individual early-type galaxies in the \hdfn\ (solid symbols),
relative to the prediction from the FP of the nearby Coma cluster.
Open circles show field galaxies from Faber et al.\ (1989) and
van Dokkum et al.\ (2001);
dots show cluster galaxies from J\o{}rgensen et al.\ (1996),
van Dokkum et al.\ (1998), and Kelson et al.\ (2000a).
All data except those of Faber et al.\ were
analyzed using the same procedures.
For reference, the lines
indicate stellar formation redshift of 3 (solid) and 1.5 (broken).
In panel {\em (b)} we show the FP of
field early-type galaxies edge-on, after correcting the surface
brightnesses for the observed luminosity evolution of cluster
galaxies. High redshift field and cluster galaxies
follow very similar trends in both panels.
\label{result.plot}}
\end{center}
\end{figure*}


\subsection{Structural Parameters}
\label{struct.sec}

For the FP analysis, effective radii $r_e$ and effective surface
brightnesses $\mu_e$ were determined by fitting 2D $r^{1/4}$
models convolved with the PSF to the $I_{814}$ images (see,
e.g., {van Dokkum} \& {Franx} 1996; {Kelson} {et~al.} 2000b). The drizzled ``Version 2''
data release of the \hdfn\ ({Williams} {et~al.} 1996) was used for
the analysis. We used four well exposed, non-saturated stars
sampling the full spatial varation of the WFPC2 field to
approximate the true PSFs. The rms range in structural parameters
is $\approx 6$\,\% in $r_e$, and $\approx 10$\,\% in $\mu_e$. The
errors are highly correlated, and 
the rms uncertainty in the product $r_e \mu_e^{0.8}$ which enters
the FP is $\approx 2$\,\%.
Median values of $r_e$ and $\mu_e$ (in units of $I_{814}$
magnitudes per square arcsecond) are listed in Table 1. To compare
galaxies at different redshifts, surface brightnesses were
converted to a common rest-frame band. For galaxies at $z=0.8-1$
the observed $I_{814}$ band closely corresponds to the rest-frame
$B$ band; hence we transformed the $I_{814}$ surface brightnesses
to rest-frame $B$ using the observed $V_{606} - I_{814}$ colors,
as explained in {van Dokkum} \& {Franx} (1996).

%

\section{Mean Ages from the Fundamental Plane}

We determine the mean age of the stellar populations in $z\simeq
1$ early-type galaxies from the evolution of the mean $M/L_B$
ratio (see {Franx} 1993). Although derivation of absolute
ages from morphologically-selected samples is susceptible to
progenitor bias ({van Dokkum} \& {Franx} 2001), we can constrain the
{\em relative} ages of field galaxies compared to those of cluster
galaxies.

As demonstrated by, e.g., Treu et al.\ (2001), the evolving
$M/L_B$ ratio can be derived from samples spanning a large
redshift range by calculating the offsets of individual galaxies
from the prediction of the locally determined FP. The nearby FP of
cluster galaxies has the form $r_e \propto \sigma^{1.20}
I_e^{-0.80}$ in the $B$ band (J\o{}rgensen, Franx, \&
Kj\ae{}rgaard 1996), which implies that $M/L \propto \sigma^{0.50}
r_e^{0.25} \propto M^{0.25}, $ assuming that early-type galaxies
form a homologous family (Faber et al.\ 1987). Further assuming
that the FP of distant field early-type galaxies has the same form
as that in nearby clusters, the residual from the FP is related to
an offset in $M/L_B$ ratio through

\begin{equation}
\Delta M/L_B \propto \sigma^{1.50} r_e^{-1.25} I_e^{-1}
\end{equation}
(see {van Dokkum} {et~al.} 2001).

The $\Delta \ln M/L_B$ ratios of the \hdfn\ galaxies are shown as a function
of redshift in Fig.\ \ref{result.plot}(a). Included are cluster
galaxies, local field galaxies, and the field sample at
$0.15<z<0.55$ from van Dokkum et al.\ (2001).
We find $\Delta \ln M/L_B \propto (-1.25 \pm 0.15) z$
for the full sample of field galaxies, compared to $\Delta \ln M/L_B
\propto (-1.12 \pm 0.11) z$ for clusters. Assuming the same rate
of evolution for both samples and using the biweight estimator
({Beers}, {Flynn}, \& {Gebhardt} 1990), we find a systematic offset between the
\hdfn\ galaxies and cluster galaxies of $-0.14 \pm 0.13$ in $\ln
M/L_B$. Assuming a {Salpeter} (1955) IMF and no systematic
difference in metallicity, this offset corresponds to an age
difference of $16 \pm 15$\,\% at $\overline{z}=0.88$. The slow evolution
of field galaxies that we measure
strengthens conclusions from earlier work by Kochanek et al.\ (2000),
 {Treu} {et~al.} (2001), and {van Dokkum} {et~al.} (2001). Treu et al.\ (2002)
find stronger evolution of
$\Delta \ln M/L_B \propto -1.68^{+0.25}_{-0.37}$ for their field sample.
This (marginally significant)
inconsistency could be due to
subtle morphological selection effects
(see \S 4), differences in the analysis, or other effects.

In Fig.\ \ref{result.plot}(b) we show the
field FP viewed edge-on, with surface brightnesses corrected
according to luminosity evolution of cluster galaxies. The FP is
remarkably similar to that measured in local clusters (solid
line). The biweight scatter in the \hdfn\ galaxies is $\approx
0.1$ in $\log r_e$. Larger samples are needed to determine the
scatter more accurately, and to determine whether there is
evolution in the tilt.

The similarity between cluster galaxies and field galaxies is
remarkable in light of expectations from hierarchical galaxy
formation models, which predict that field galaxies should be
substantially younger than cluster galaxies. The {Diaferio} {et~al.} (2001)
models predict a difference of $\Delta \ln M/L_B \approx
-0.6$ (see van Dokkum et al 2001), which is ruled out by the data.

\section{Evidence for Recent Mass Growth}

Two galaxies deviate significantly from the general trends in
Fig.\ \ref{result.plot}, suggesting they have young stellar
populations. The object with the lowest $M/L_B$ ratio (17) is the
only galaxy in our sample with a blue core
({Menanteau} {et~al.} 2001), which is likely associated with an
active nucleus\footnote{Curiously our spectrum shows no evidence
for nuclear activity.}
(Hornschemeier et al.\ 2000, Sarajedini
et al.\ 2000).
In Fig.\ \ref{res.plot} we show the
ultra-deep WFPC2 images of these galaxies, after subtracting the
best fitting $r^{1/4}$ laws.  Remarkably, both deviating objects
have morphological signatures indicative of recent accretion or
mergers. These features have surface brightness $\mu_{814} =
24.5-25$, and are detectable only because
of the depth of the HDF images. None of the other seven
galaxies show unambiguous deviations from axisymmetry. We conclude
that there is tantalizing evidence for recent mass growth in the
population of $z\simeq 1$ early-type galaxies.
Note that these
anomalous objects were {\em not} excluded from the field vs.\ cluster
comparison in \S\ 3, although the $M/L_B$ ratios of
these galaxies are highly uncertain (in particular for galaxy
17).

\null
\vbox{
\begin{center}
\leavevmode
\hbox{%
\epsfxsize=7.8cm
\epsffile{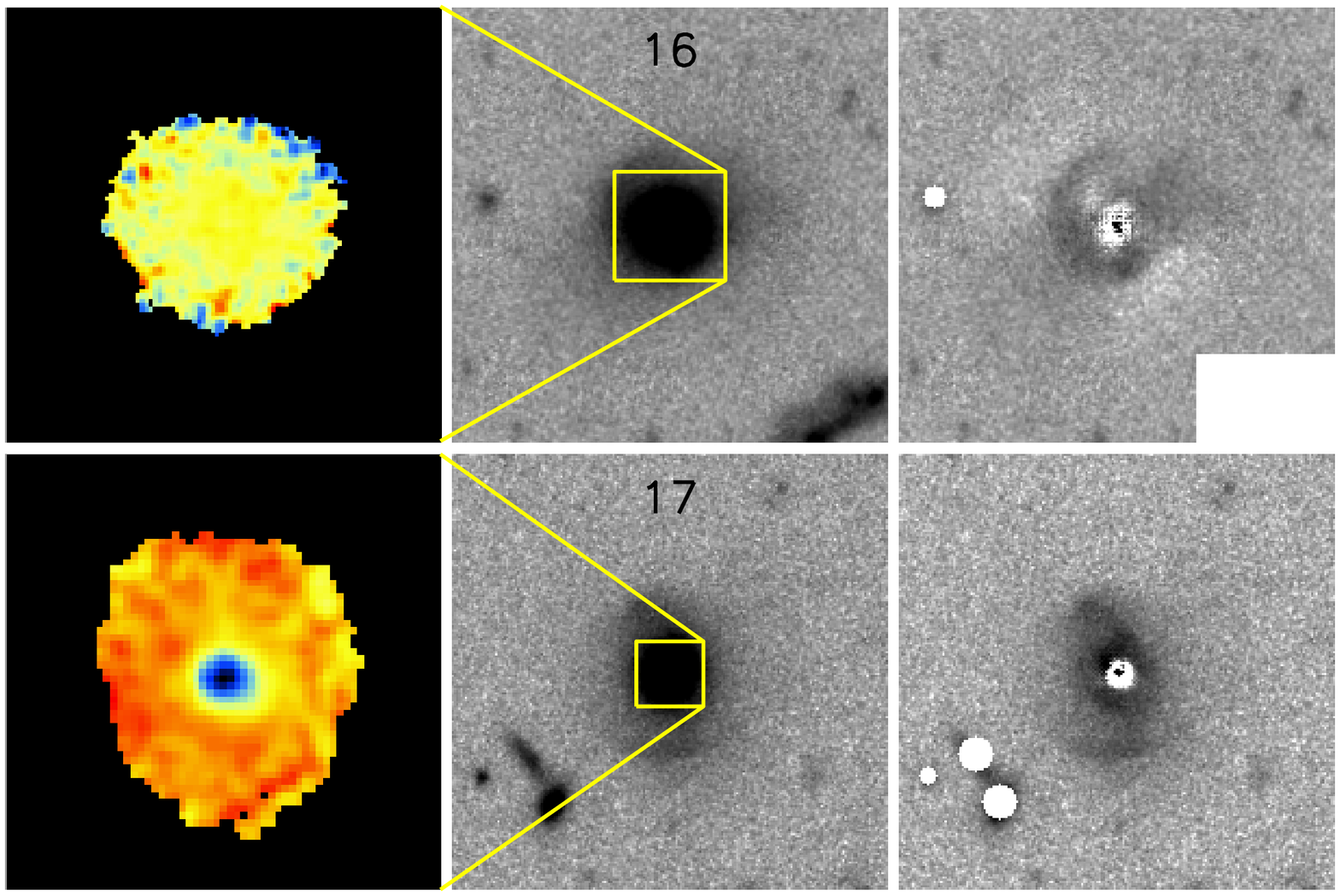}}
\figcaption{\small
Residual images for two deviant objects in the FP,
obtained by subtracting the best fitting $r^{1/4}$ law
profiles (right panels). Both show clear non-axisymmetric residual
structure, probably of tidal origin. 
The left panels show the pixel-by-pixel
color variation. These
faint structures could only have been reliably detected in the
ultra-deep HDF data.
\label{res.plot}}
\end{center}}

\section{Evidence for Recent Star Formation}

Whereas $M/L_B$ ratios are sensitive to the luminosity weighted
mean age of stellar populations, age-sensitive spectral
diagnostics such as the H$\delta$ absorption line provide
information on the presence of (small amounts of) young stars
(e.g., {Barger} {et~al.} 1996). In Fig.\ \ref{coadd.plot} we show the
mean spectrum of the ten early-type galaxies in our sample,
created by normalizing each spectrum at $\lambda_{\rm
rest}=4050$\,\AA. The median redshift $\overline{z}=0.90$, and the
biweight average is $\overline{z}_{\rm BI}=0.86$. The S/N is
non-uniform as the full wavelength region is not covered by every
galaxy; it is highest ($\sim 40$)
in the region near the \hk\ break.
Shown for comparison is the coadded spectrum of 356 red
galaxies at $0.30<z<0.35$ in the Sloan Digital Sky Survey
(SDSS; {Eisenstein} {et~al.} 2003).

The mean spectrum clearly reveals prominent Balmer absorption
lines in addition to weak \3727\ emission, in striking contrast
to the color-selected, lower redshift
SDSS spectrum. The equivalent widths
are $W_{\lambda}({\rm H}\delta)=4.0^{+0.9}_{-0.5}$ and
$W_{\lambda}$(\oii)$=-1.7^{+0.5}_{-1.1}$, using bandpass
definitions of {Fisher} {et~al.} (1998). Uncertainties were derived by
bootstrap resampling; we note that Balmer absorption lines at
levels consistent with the average value can be detected in almost
all individual spectra, albeit with lower significance (errors
$\approx 1.0$\,\AA; Table 1).

\null
\vbox{
\begin{center}
\leavevmode
\hbox{%
\epsfxsize=7.8cm
\epsffile{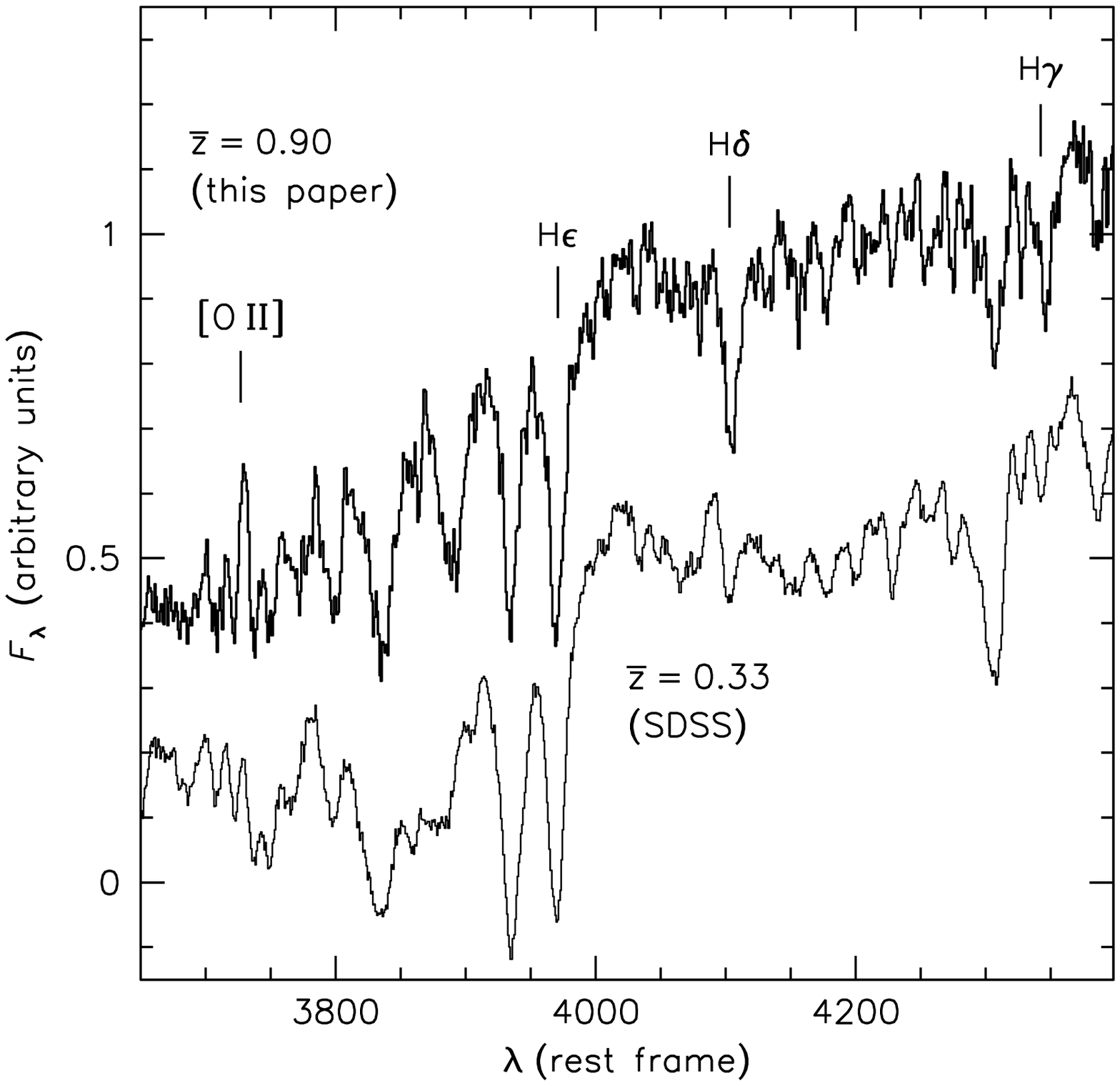}}
\figcaption{\small
Coadded rest-frame spectrum for 10 \hdfn\
early-type galaxies at $\overline{z}=0.90$. The Balmer absorption
lines are clearly enhanced
with respect to the (color-selected)
$\overline{z}=0.33$ mean SDSS spectrum
(Eisenstein et al.\ 2003). 
The SDSS spectrum is offset by 0.35 in $F_{\lambda}$.
\label{coadd.plot} }
\end{center}}

\section{Discussion}

Our small sample of \hdfn\ galaxies tells a complex tale:
the similarity of the $M/L_B$ ratios with those of cluster galaxies
on one hand contrasts with evidence for
recent accretion and star formation on the other. The Balmer
absorption strength of $4.0$\,\AA\ is remarkably high, and exceeds
the expectation from simple passive evolution of a single age
population: taking $2<z_{\rm form}<3$ as the formation epoch of
stars in cluster and field galaxies ({van Dokkum} \& {Franx} 2001, and \S\
3), the {Vazdekis} (1999) models predict
$W_{\lambda}({\rm H}\delta)\approx 1.0$\,\AA\ for galaxies viewed
at $z\simeq 0.9$.

Our results can be explained consistently by postulating that a
substantial fraction of the \hdfn\ galaxies experienced some
recent star formation, but that the bulk of the stellar mass was
formed at redshifts $z\gtrsim 2$. We used the {Vazdekis} (1999)
models to generate composite spectra consisting of a pre-existing,
2.8\,Gyr old population containing the bulk of the stellar mass
and a low mass young component of varying age. Such models
naturally produce strong effects in the H$\delta$ line and only
modest changes in $\log M/L_B$. As an example, a 0.5\,Gyr old
secondary star burst containing $3$\,\% of the total stellar mass
gives $W_{\lambda}({\rm H}\delta)= 3.5$\,\AA\ and $\Delta \ln
M/L_B = -0.18$, entirely consistent with the observations.
We note that similar models have been proposed for nearby
early-type galaxies by
Trager et al.\ (2000).

It is difficult to unambiguously identify the physical origin of
recent star formation in the \hdfn\ galaxies. Tidal features
detected in two galaxies argue that at least in some cases the
cause may be a merger or the capture of a satellite. The small
mass associated with the young component argues against major
mergers of gas-rich disk systems in the redshift interval
$1.0\lesssim z \lesssim 1.5$. Mergers between bulge-dominated,
gas-poor systems are much more difficult to constrain, as they
would not have a large effect on the $M/L_B$ ratios and do not
develop well defined tidal tails (e.g., {Barnes} 1988). Such
mergers have been observed in clusters (e.g., {van Dokkum} {et~al.} 1999),
and were recently invoked to explain the inferred evolution of the
luminosity density of red field galaxies (Bell et al.\ 2003).

Of course, we cannot yet exclude the possibility that the
small \hdfn\
sample is special. The stacked spectrum of 15 ``old'' Extremely
Red Objects (ERO) presented by {Cimatti} {et~al.} (2002) shows no
evidence for enhanced Balmer absorption, suggesting significant
variation in the properties of $z\simeq 1$ early-type galaxies
depending perhaps on how they are selected.
Only by correlating morphologies, $M/L$ ratios,
colors and spectral features of large samples is it
likely that a quantitative assembly history for $z\simeq 1$
early-type galaxies can be determined.

\begin{acknowledgements}

We thank Felipe Menanteau and Bob Abraham for their significant
contributions in the early stages of this project and acknowledge
discussions with Joshua Barnes, Andrew Benson, Jarle
Brinchmann, Alan Dressler, Jim Gunn and Tommaso Treu.
We thank the anonymous referee for constructive
comments which improved the manuscript.
This work was partly supported by STScI grant HST-AR-09541.01-A.
The authors recognize and acknowledge the
cultural role and reverance that the summit of Mauna Kea has
always had within the indigenous Hawaiian community. We are most
fortunate to have the opportunity to conduct observations from
this mountain.

\end{acknowledgements}


\end{document}